\documentclass{elsart}
\usepackage{exscale}                  
\usepackage[intlimits]{amsmath}       
\usepackage{amsfonts}
\usepackage{amssymb,amscd}
\usepackage[dvips]{epsfig}                   
\usepackage{array}
\usepackage{wrapfig}
\newcommand{\beqa}{\begin{eqnarray}}
\newcommand{\eeqa}{\end{eqnarray}}
\newcommand{\beq}{\begin{equation}}
\newcommand{\eeq}{\end{equation}}

\begin{document}

\begin{frontmatter}

\title{
Vertex functions and infrared fixed point in Landau gauge SU(N)
Yang-Mills theory}
\vspace{-10mm}
\author{R.~Alkofer$ ^1$, C.~S.~Fischer$ ^2$,
F.~J.~Llanes-Estrada$ ^3$}
\address{$ ^1$Institute for Theoretical Physics, U.\ T\"ubingen, 
D-72076 T\"ubingen, Germany \\
$ ^2$IPPP, University of Durham, Durham DH1 3LE, U.K. \\
$ ^3$Dept. F\'{\i}sica Teorica I, Univ. Complutense, Madrid
28040, Spain}

\date{30 December 2004}
\begin{abstract}
The infrared behaviour of vertex functions in an $SU(N)$ Yang-Mills  theory in
Landau gauge is investigated employing a skeleton expansion of the
Dyson-Schwinger equations. The three- and four-gluon vertices become singular
if and only if all external momenta vanish while the dressing of the
ghost-gluon vertex remains finite in this limit. The running coupling as
extracted from either of these vertex functions possesses an infrared fixed
point. In general, diagrams including ghost-loops dominate in the infrared
over purely gluonic ones.
\end{abstract}
\end{frontmatter}


Fifty years after the formulation of Yang-Mills theory its infrared (IR)
structure is still largely  unknown despite the fact that this knowledge is
central to any effort in understanding the strong interactions from first
principles. 
It has long been conjectured that IR enhancements are present. Indeed they are  
necessary to explain confinement. Such IR enhancements may also be the reason
that no explicit glue becomes visible in the low-lying hadron mass spectrum. 

Lattice calculations include in principle all non-perturbative 
features of Yang-Mills theories but are in practice limited  
by the finite lattice volume in the study of possible IR singularities
\cite{Bonnet:2001uh,Langfeld:2001cz,Bowman:2004jm,Silva:2004bv}.
A complementary non-perturbative continuum method is provided by the
Dyson-Schwinger equations (DSEs). In Landau gauge 
the DSEs for the ghost and gluon propagators have been
analytically solved in the  IR assuming ghost dominance
\cite{Alkofer:2001wg,vonSmekal:1997is,Atkinson:1998tu,Zwanziger:2001kw,Lerche:2002ep,Fischer:2002hn,Alkofer:2003jj}. 
Lattice and DSEs are complementary  and yet they agree on 
the propagators' IR behaviour: there is clear evidence for an 
IR finite or even vanishing gluon propagator and a strongly
diverging ghost propagator, in accordance with both, the 
Kugo-Ojima confinement criterion \cite{Kugo:1995km} and the 
Gribov-Zwanziger scenario \cite{Zwanziger:2003cf}.

The running of the gauge coupling is intimately related to the momentum
dependence of the primitively divergent vertex functions in an $SU(N_c)$
Yang-Mills theory. For sufficiently large momenta the coupling can be 
calculated from perturbative corrections to these vertex functions
\cite{Celmaster:1979km}. The breakdown of perturbation theory is signaled
by Landau poles, unphysical singularities at non-vanishing space-like
momenta. By simply imposing analyticity for space-like momenta 
extrapolations to the IR have been performed in so-called Analytic Perturbation
Theory  \cite{Shirkov:1997wi,Solovtsov:1999in,Howe:2003mp}. 
Typically these studies find a well-behaved coupling at all 
momenta and an IR fixed point. Qualitatively this agrees with 
the findings from the genuinely non-perturbative continuum approaches. 
In the latter the IR behavior of the ghost-gluon 
interaction in Landau gauge has been determined either from DSEs 
or the Exact Renormalization Group Equations (ERGEs)
\cite{Zwanziger:2001kw,Lerche:2002ep,Pawlowski:2003hq,Fischer:2004uk} 
and yield an IR fixed point with $\alpha(0) \approx 8.9/N_c$. 
The corresponding couplings from the three- and four-gluon vertex functions
have not yet been studied with these techniques.
Within perturbation theory the principle of gauge invariance leads to the
universality of the gauge coupling: All three couplings are 
equal up to the well-known renormalization scheme dependencies
\cite{Celmaster:1979km,Stevenson:1981vj,Brodsky:1982gc} at every finite order.
Although a general proof is lacking one expects such relations to also hold in
the full non-perturbative theory.

In this letter we propose a method to investigate the IR behaviour of Greens
functions with an arbitrary number of ghosts and gluons.  We detail the results
for the gluon-self interaction vertices thus completing our knowledge about the
primitively divergent vertex functions ({\it i.e.\/} those appearing in the  
renormalized Lagrangian) and the corresponding running coupling(s) in the deep
IR. We construct a skeleton expansion for each DSE employing the fully dressed
primitively divergent $n$-point functions of Yang-Mills theory. To all orders
in this  expansion the three- and four-gluon vertex  functions are IR singular 
if and only if all  external momenta vanish. The corresponding IR
exponents are hereby proportional to the IR exponent of the ghost
renormalization function such that the corresponding couplings have IR fixed
points. The results presented in this letter verify the self-consistency of
ghost dominance in a quite general way and prove that ghost dominance
is a successful guiding principle when determining the IR behaviour
of Yang-Mills Greens functions in the confining phase.

In Landau gauge, the ghost and gluon propagators in Euclidean momentum space 
are described by the renormalization functions $G(p^2)$ and $Z(p^2)$:
\beq
D^G(p^2) = - \frac{G(p^2)}{p^2} \, , \qquad
D_{\mu \nu}(p^2)  = \left(\delta_{\mu \nu} -\frac{p_\mu 
p_\nu}{p^2}\right) \frac{Z(p^2)}{p^2} \, .
\eeq
On the other hand, the three- and four-point functions feature tensor 
structures not present in the bare  Lagrangian thus requiring multiple such  
scalar functions. We will not specify these yet and discuss first the  behaviour of
the scalar amplitude multiplying only the tensor defined on the level of
the bare Lagrangian. {\it E.g.\/} for the three-gluon vertex this tensor reads
\beq
\delta_{\mu_1 \mu_2}(p_1-p_2)_{\mu_3} +
\delta_{\mu_2 \mu_3}(p_2-p_3)_{\mu_1} +
\delta_{\mu_3 \mu_1}(p_3-p_1)_{\mu_2} \, .
\eeq
In a setting where all external momenta $(p_i)^2$ vanish as a single momentum scale
$p^2\to 0$ this tensor will be multiplied by an amplitude
$H^{3g}_1(p^2)$ in the fully dressed three-gluon vertex.
For the  four-gluon and ghost-gluon vertex functions
the amplitudes $H^{4g}_1(p^2)$ and $H^{gh-g}_1(p^2)$ are defined analogously.
Starting from the established result 
\cite{Zwanziger:2001kw,Lerche:2002ep,Pawlowski:2003hq,Fischer:2004uk}
\beq
Z(p^2) \to (p^2)^{2\kappa}  \, ,\qquad
G(p^2) \to (p^2)^{-\kappa}  \, , \label{Zbeh}  
\eeq
we will show in the following that for $p^2\to 0$
\beq
H^{3g}_1(p^2) \to (p^2)^{-3\kappa} \, , \qquad
H^{4g}_1(p^2) \to (p^2)^{-4\kappa} \, , \qquad 
H^{gh-g}_1(p^2) \to {\rm const.} 
\label{Hgbeh} 
\eeq
where the parameter $\kappa$ is determined from the propagator DSEs,
see below.  
Eqs.\ (\ref{Zbeh}) imply that the minimum of a gluon's dispersion relation 
does not occur at zero momentum.\footnote{The phenomenological consequences of 
an IR suppressed gluon propagator can for example be infered from ref.\
\cite{cotanch}.} In addition, it concurs with Zwanziger's picture
\cite{Zwanziger:2003cf}: the geometric degrees of freedom dominate IR
Yang-Mills theory. 
Eqs.\ (\ref{Hgbeh}), on the other hand, state that the self-interactions of 
gluons become large at very low momentum. The connected three- and four-gluon
functions hereby violate the bound conjectured in ref.\ \cite{Zwanziger:1990by},
although only at one specific kinematical point.

\begin{figure}[t]
\centerline{\epsfig{file=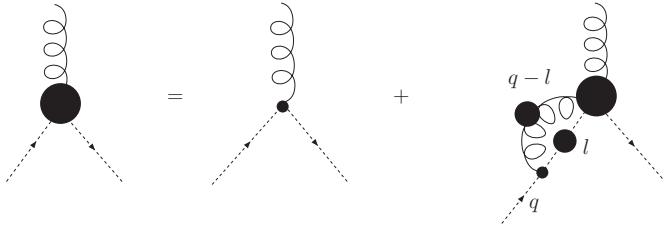,width=90mm}}
\caption{Ghost-gluon vertex DSE.}
\label{DSE-ghg}
\end{figure}

It will be demonstrated in the following that each of the eqs.\ (\ref{Zbeh})
and (\ref{Hgbeh}) is correct when the others are assumed as given. At the end,
one thus obtains a self-consistent solution of the DSE system in the IR. The
starting point is provided by the observation that in Landau gauge  the
ghost-gluon vertex is UV finite and remains bare for vanishing incoming ghost
momentum  \cite{Taylor:ff,Marciano:1977su}. This is easily shown using the DSE
for  the full ghost-gluon vertex, see fig.\ \ref{DSE-ghg}. The bare ghost-gluon
vertex in the interaction diagram is proportional to the internal loop momenum
$l_\mu$. Due to the transversality of the gluon propagator  $D_{\mu \nu}$ one
has  $l_\mu D_{\mu \nu}(l-q) = q_\mu D_{\mu \nu}(l-q)$. Thus,
the interaction diagram vanishes for $q_\mu \rightarrow 0$. This argument would
only be invalidated if the two-ghost--two-gluon scattering kernel had an IR
divergence. However, recent lattice results and numerical DSE studies 
of the ghost-gluon vertex \cite{Schleifenbaum:2004id,Cucchieri:2004sq} agree 
with a bare vertex in the IR. From our analytical study it also follows 
that such a divergence is absent. 

\begin{figure}[t]
\centerline{\epsfig{file=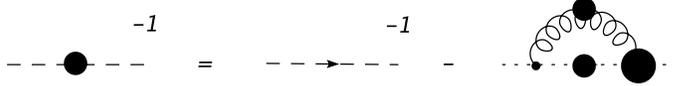,width=90mm}}
\caption{Ghost propagator DSE.}
\label{DSE-gh}
\end{figure}

The IR behaviour of two-point functions has been adressed within ERGEs and DSEs
\cite{vonSmekal:1997is,Atkinson:1998tu,Zwanziger:2001kw,Lerche:2002ep,Fischer:2002hn,Pawlowski:2003hq,Fischer:2004uk}.
We briefly discuss the IR aspects of the ghost propagator DSE depicted in fig.\
\ref{DSE-gh}.
With an IR finite ghost-gluon vertex, see the last of eqs.\ 
(\ref{Hgbeh}), and a power law ansatz for the dressing functions at 
low external $p^2$,
$Z(p^2) = A\, (p^2)^{\alpha}$ and $G(p^2) = B\, (p^2)^{\beta}$,
a self-consistent solution  arises with
$\kappa:=-\beta=2\alpha$ from matching the exponents of
both sides of the ghost DSE \cite{Watson:2001yv}. 
Hereby the formula for a scalar
integral in $d$ dimensions, depending on one external momentum 
$p_\mu$
\begin{equation} \label{IR-int}
\int d^dq (q^2)^a ((q-p)^2)^b =
(p^2)^{d/2+a+b} \frac{\Gamma(d/2+a) \Gamma(d/2+b)
\Gamma(-a-b-d/2)}{\pi^2 \Gamma(-a) \Gamma(-b) \Gamma(d+a+b)} 
\end{equation}
has been used.
The actual value of $\kappa$ follows from IR consistency of the ghost with
the gluon propagator DSE, and depends slightly on the details of the 
ghost-gluon vertex' (finite) dressing \cite{Lerche:2002ep}. For a bare vertex 
one obtains $\kappa = (93 - \sqrt{1201} )/98  \approx 0.595 $.
Note, however, that the precise value of $\kappa$ will be irrelevant for
all arguments presented in the following as long as $0 < \kappa < 1$  
\cite{Lerche:2002ep,Watson:2001yv} and $\kappa \ne 1/2$  \cite{Lerche:2002ep}.

\begin{figure}[t]
\centerline{\epsfig{file=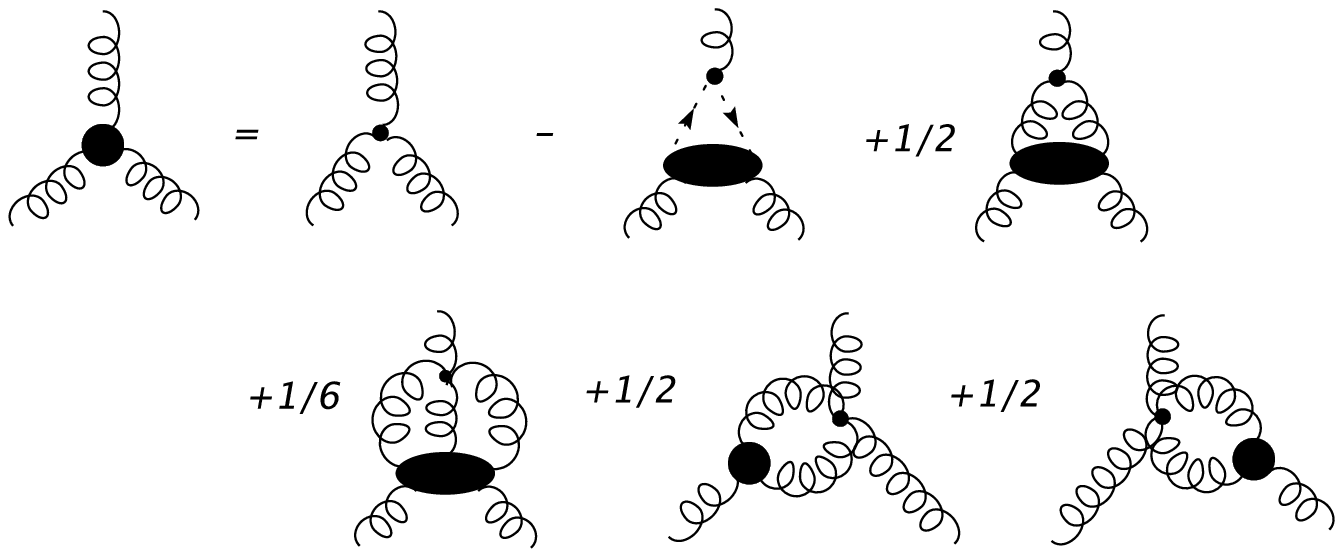,width=105mm}}
\vspace*{8mm}
\centerline{\epsfig{file=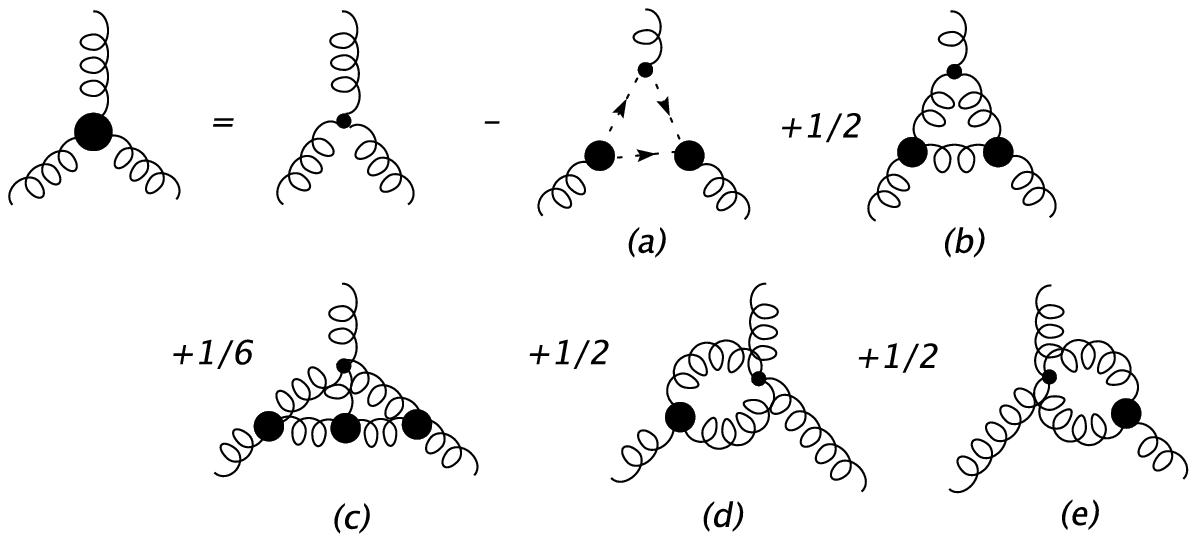,width=105mm}}
\caption{Exact Dyson-Schwinger equation for the three-gluon vertex
and  lowest order in a skeleton expansion of the four-
and five-point functions. All internal propagators in the diagrams are to
be understood as fully dressed.} 
\label{DSE-3g}
\end{figure}
\begin{figure}[t]
\centerline{\epsfig{file=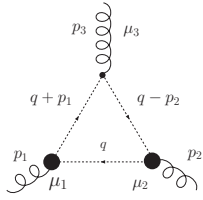,width=30mm}}
\caption{\label{DSE-3g-kin}
Momentum routing in the ghost-loop diagram of the three-gluon
vertex. All external momenta flow into the loop. Internal propagators are fully dressed.}
\end{figure}

In a first step we establish the three-gluon vertex power law in
eq.\ (\ref{Hgbeh}).
The corresponding DSE (see fig.\ \ref{DSE-3g}) includes four- and five-point
functions which we treat in a skeleton expansion in terms of the fully dressed,
primitively divergent Greens functions. Furthermore, we develop counting rules
for the IR degree of divergence of a diagram. To lowest order in the expansion
we can identify each  subdiagram with its leading IR singularity. 
We will demonstrate that all higher order diagrams are either equally or less
IR singular. If the three momenta entering the vertex  are much smaller than
$\Lambda_{\tt QCD}$ the integral will be dominated by loop momenta also smaller 
than $\Lambda_{\tt QCD}$ due to the denominators of the propagators. 
The dressing functions of these propagators are then the simple power laws 
given in eqs.\ (\ref{Zbeh}), allowing to integrate the loops 
analytically. 
The ghost-loop diagram $(a)$ in fig.\ \ref{DSE-3g} is IR leading. We thus 
evaluate this diagram first and then substitute the resulting
power law for the three-gluon vertex into the other diagrams to check for
self-consistency.  This is simplest demonstrated at the symmetric kinematical point
\begin{equation}
p_1^2 = p_2^2 = p_3^2 = p^2, \qquad 
(p_1\cdot p_2) = (p_1\cdot p_3) = (p_2\cdot p_3) \equiv -\frac{1}{2}p^2 .
\end{equation}
The tensor structure of the three-gluon vertex then reduces from fourteen tensors 
at general kinematics to three independent ones \cite{Celmaster:1979km}. 
Omitting the overall color factor and the delta function expressing 
momentum conservation the vertex is then given by
\beqa
\Gamma_{\mu_1 \mu_2 \mu_3}(p_1,p_2,p_3) &=& H_1^{3g}(p^2) \,
\left( \delta_{\mu_1 \mu_2}(p_1-p_2)_{\mu_3} + {\rm cyclic \, permutations} 
\right) \nonumber\\
%
 &-& H_2^{3g}(p^2)/p^2 \,\, (p_2-p_3)_{\mu_1}(p_3-p_1)_{\mu_2}(p_1-p_2)_{\mu_3}\nonumber\\
 &+& H_3^{3g}(p^2)/p^2 \, \left( p_{1_{\mu_3}} p_{2_{\mu_1}}p_{3_{\mu_2}} - 
p_{1_{\mu_2}} p_{2_{\mu_3}}p_{3_{\mu_1}} \right) \label{3g-symmvert}
\eeqa
with the momentum routing and indices as given in fig.\ \ref{DSE-3g-kin}.

We are interested in the limit $p^2 \rightarrow 0$ where the ghost-gluon vertex
becomes bare again \cite{Taylor:ff}, $\Gamma^{gh}_{\mu}(p,q) = i q_\mu $ with
$q_\mu$ being the momentum of the outgoing ghost. Substituting the power laws,
eqs.\ (\ref{Zbeh}), for the ghost and gluon  dressing functions and
employing (\ref{IR-int}) the IR behavior of the dressing functions $H_1^{3g}$, $H_2^{3g}$
and $H_3^{3g}$ of the vertex is determined analytically to be
\begin{eqnarray} \label{ghost1}
H_1^{3g}(p^2) &=&  - h(\kappa) \; 
{\displaystyle \frac {1}{36}} \;  (\phantom{1}97\kappa^{2} - 205 \kappa +
100) \; \;  (p^2)^{(-3  \kappa)}\\ \label{ghost2}
H_2^{3g}(p^2) &=& \phantom{-} h(\kappa) \; 
 {\displaystyle \frac {2}{27}}\;  (\phantom{1}59\kappa^{2} - 131 \kappa + \phantom{1}56) 
 \; \; (p^2)^{(-3 \kappa)}\\ \label{ghost3}
H_3^{3g}(p^2) &=& \phantom{-}h(\kappa) \; 
 {\displaystyle \frac {1}{18}} \; (119\kappa^{2} - 323 \kappa +
164) \; \; (p^2)^{(-3 \kappa)}
\end{eqnarray}
with
\begin{equation}
h(\kappa) = \frac{-g^2 N_c B^3}{32 \pi^2}
\frac{\Gamma(3\kappa) \Gamma(1-2\kappa) \Gamma(1-\kappa)}{\Gamma 
(1+\kappa)\Gamma(2+2\kappa)\Gamma(3-3\kappa)}
\end{equation}
and $B$ being the leading IR coefficient of the ghost.
Crucial to these expressions are the momentum power laws.
All three dressing functions are proportional  to $(p^2)^{(-3\kappa)}$,
{\it i.e.\/} with  $\kappa > 0$ \cite{Watson:2001yv}  the 
three-gluon vertex is IR singular, and the first of eqs. (\ref{Hgbeh}) 
is established.
Note that the degree of singularity is the same for all three dressing functions, 
and given by three times $-\kappa$, the latter being the IR exponent of the 
ghost propagator. In the following it will
become evident that the IR anomalous dimension of a diagram is simply the sum of
IR exponents of propagators and vertices constituting the diagram. 
All trivial dimensions of the vertex are due to the tensor structure
whereas the dressing functions $H_1^{3g}, H_2^{3g}$ and $H_3^{3g}$ are dimensionless. Any
anomalous dimension appearing 
in the diagram of fig. \ref{DSE-3g-kin} via nontrivial vertex or propagator dressings 
therefore has to appear in the functions $H_1^{3g}, H_2^{3g}$ and $H_3^{3g}$. Therefore
we can determine the degree of divergence of Greens functions
just by counting IR anomalous dimensions.
As $p^2$ is the only scale, 
$(p^2)^\rho$ with $\rho=-3\kappa$ has necessarily to arise in this case.

Note that a dressed ghost-gluon vertex would not have changed the running
of $H_1^{3g}, H_2^{3g}$ and $H_3^{3g}$ with momentum, since in
Landau gauge it carries no overall anomalous dimension. IR
corrections to the vertex dressings can therefore 
be typically expressed as quotients $p^n/q^n$ \cite{Lerche:2002ep}. Such 
dressings cannot change the momentum dependence in eqs.
(\ref{ghost1}-\ref{ghost3}).
The following observation is important: The diagram in fig.\
\ref{DSE-3g-kin} becomes
IR singular if and only if all three incoming momenta $(p_1)^2, (p_2)^2$
and $(p_3)^2$ approach zero.
This can be proven in general kinematics.\footnote{
Using appropriate
projectors to single out the scalar dressing functions of the vertex one
always ends up with scalar `massless' triangle integrals with nontrivial
powers of internal momenta. These integrals can be performed along the methods  
described in refs.\ \cite{Boos:1990rg,Anastasiou:1999ui}. 
As we have checked explicitly, no IR singularities arise unless all three
squared momenta vanish.} 
A convenient kinematic section to see this directly is {\it e.g.\/}
provided by
$p_1^2={\rm constant}$, $p_2=b p_1$ with $b$ a constant that we will 
allow to approach zero. Due to momentum conservation 
the three momenta are then collinear and only one 
of them is small. With a convenient choice for the tensor basis, the 
leading coefficient calculated from the ghost triangle graph is
\begin{eqnarray}
&H^{3g}_1 = \frac{g^2 N_c}{2}
(\tilde{Z}_1)^3 \int \frac{d^4k }{(2\pi)^4}
\frac{b(b-3)(k^2p_1^2-k\cdot p_1^2)+(b-1)2k^2 k\cdot p_1}
{6p_1^2 (b^2-3b+3)} \nonumber \\
& \hspace{40mm} \frac{G(k^2)}{k^2}
\frac{G((k+bp_1)^2)}{(k+bp_1)^2} \frac{G((k-p_1)^2)}{(k-p_1)^2}  \; .
\end{eqnarray}
In the limit $b\to 0$ the integrand is singular but integrable, and thus 
the corresponding diagram yields a finite contribution.

In the DSE for the three-gluon vertex no other term in
the skeleton expansion has a larger power IR divergence. To show this 
we count powers for the remaining diagrams in fig.\ \ref{DSE-3g}
and compare the result with eqs. (\ref{ghost1}-\ref{ghost3}). Employing
eqs.\ (\ref{Zbeh}-\ref{Hgbeh}) and denoting
the remaining four diagrams as $(b)-(e)$, {\it c.f.\/} fig.\ 3, 
we obtain their IR anomalous dimensions by summing up the anomalous dimensions
of all propagators and vertex functions of a diagram. Hereby a gluon propagator
contributes an anomalous dimension of $2\kappa$, a ghost propagator one of $(-\kappa)$,
the three-gluon vertex one of $(-3\kappa)$ and the ghost-gluon vertex carrys zero
anomalous dimensions, {\it c.f.\/} eqs.\ (\ref{Zbeh}) and (\ref{Hgbeh}). 
This yields for their respective IR exponents $\rho$:
\vspace*{-5mm}
\beqa
(b)& \; \rho = 3\times 2\kappa + 2 \times  (-3\kappa) = 0, \quad 
&(c) \; \rho = 5\times 2\kappa + 3 \times  (-3\kappa) = \kappa , \nonumber \\ 
(d)& \; \rho = 2\times 2\kappa + 1 \times  (-3\kappa) = \kappa , \quad 
&(e) \; \rho = 2\times 2\kappa + 1 \times  (-3\kappa) = \kappa .
\eeqa
Thus they are all subleading if compared to the ghost triangle diagram (a).

\begin{figure}[t]
\centerline{\epsfig{file=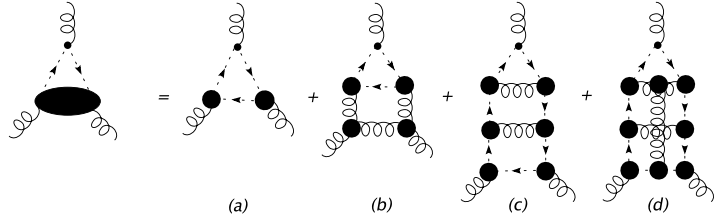,width=100mm}}
\caption{\label{skelkorr1}
Typical terms of the skeleton expansion of the ghost-gluon
scattering kernel in the DSE for the three-gluon vertex.
Internal propagators are fully dressed.} 
\end{figure}
\begin{figure}[t]
\centerline{\epsfig{file=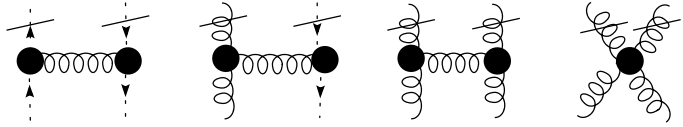,width=90mm}}
\caption{\label{skelkorr2}
 Insertions generating the loop expansion in fig.\ \ref{skelkorr1}.
One should also allow for the possibility of upgrading a three to a 
four gluon vertex.}
\end{figure}
Finally we investigate further terms from the skeleton expansion. 
Typical corrections to the ghost-triangle diagram are given in fig.\
\ref{skelkorr1}. The only scale appearing in the symmetric momentum
kinematics is $(p^2)$, thus we can determine the degree of divergence of
the diagrams by counting anomalous dimensions. Denoting the higher order 
diagrams by $(b)-(d)$ we obtain for their IR anomalous dimensions
\vspace*{-3mm}
\beqa
&(b)& \; \rho = \left(3 \cdot 2 + 3 \cdot (-1) + 2 \cdot (-3) \right)\kappa = 
-3\kappa ,\quad \nonumber \\
%
%
&(c)& \; \rho = \left(2 \cdot 2 + 7 \cdot (-1) \right) \kappa =
-3\kappa , \nonumber \\
&(d)& \; \rho = \left(4 \cdot 2 + 8 \cdot (-1) + 1 \cdot (-3) \right)\kappa =
-3\kappa ,
\eeqa
with no contribution from the ghost-gluon vertex, as discussed above.
Thus all these diagrams contribute to the
same IR order as the ghost-loop diagram~$(a)$. 

Arbitrary diagrams in the expansion can be generated with 
the four elements in fig.\ \ref{skelkorr2}, by repeated
insertion into a given lowest-order diagram. 
According to our counting rules based on eqs. (\ref{Zbeh}) and (\ref{Hgbeh})
these insertions carry an overall zero IR anomalous dimension. 
Therefore in general, starting from a given diagram
in the skeleton expansion with an IR anomalous dimension $\rho$ all possible
insertions generate diagrams of the next order in the skeleton expansion
with the same degree of divergence. By induction this is true for the
whole series of skeleton diagrams up to any order.
{\em Therefore the degree of IR divergence of a fully
dressed Greens function in the presence of only one external scale can
be already read off by counting the anomalous dimensions of the lowest order
diagrams in the skeleton expansion.}  This general rule especially implies that
higher order corrections to the diagrams $(b)-(e)$ of fig.\ \ref{DSE-3g}
are all IR subleading. Thus the IR exponent 
$\rho = -3\kappa$ for the three-gluon
vertex is established.\footnote{This result has been anticipated in \cite{Watson:2000}
employing the STI $Z_1(\mu) = Z_3(\mu)/\tilde{Z}_3(\mu)$. In the presence of only one
momentum scale all renormalized dressing functions run with $(p^2/\mu^2)^\alpha$, where
$\alpha$ is the corresponding anomalous dimension.
Therefore the IR powers in Eq.(\ref{Zbeh}) imply that $Z_1(\mu) \sim (\mu^2)^{3\kappa}$
and $\rho = -3\kappa$ follows directly.
}

\vspace*{3mm}
\begin{figure}[h]
\centerline{\epsfig{file=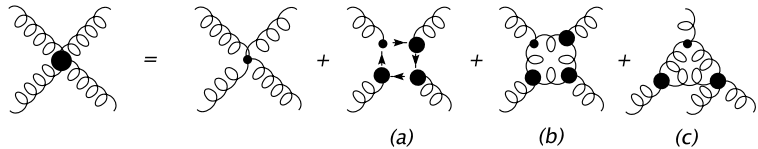,width=100mm}}
\caption{\label{4g} First order expansion of the DSE for
the four-gluon correlation. Internal propagators are to be understood as fully dressed.} 
\end{figure}

The lowest order terms in the skeleton expansion of the DSE for the
four-gluon vertex are given in fig. \ref{4g}. 
Here we consider the special momentum configuration where all four
momenta flowing into the loop are equal in magnitude and pairwise
parallel or antiparallel. Denoting the ingoing momenta
with $p_{i=1\ldots 4}$ we  have $p_1^2 = p_2^2 = p_3^2 = p_4^2 \equiv p^2$
and all three Mandelstam variables $s = (p_1+p_2)^2$, $t = (p_2+p_3)^2$
and $u=(p_1+p_3)^2$ are either zero or proportional to $p^2$. With the 
IR counting rules established above we can then determine the
degree of divergence of the three one-loop diagrams 
(a), (b) and (c). We obtain
\vspace*{-4mm}
\beqa
&(a)& \; \rho = -4\kappa , \quad\quad
(b) \; \rho = \left( 4 \cdot 2 + 3 \cdot (-3)\right)\kappa = -\kappa,\nonumber\\ 
&(c)& \; \rho = \left(3 \cdot  2 + 1 \cdot (-3) + 1 \cdot (-4) \right) \kappa
            = -\kappa .
\eeqa
Here the ghost box diagram is IR dominating, and yields $-4\kappa$
as IR exponent for the four-gluon vertex.

\begin{figure}[t]
\centerline{
\centerline{\epsfig{file=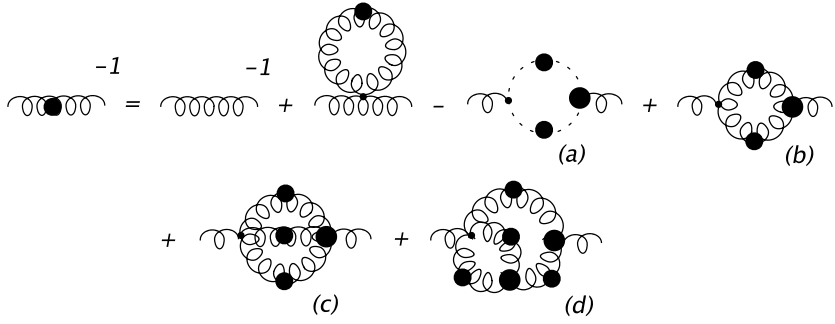,width=110mm}}}
\caption{\label{glue}
The Dyson-Schwinger equation for the gluon propagator.} 
\end{figure}

Finally, we have to check whether eqs.\ (\ref{Zbeh}) remain valid assuming 
the new results for the gluon-selfinteraction vertices.
We therefore reconsider the DSE for the gluon propagator
given diagramatically in fig.\ \ref{glue}. Again due to the denominators
of the propagators all diagrams are dominated by loop momenta similar to
the external scale $p^2$. We can thus determine the IR behaviour for
each diagram by the counting rules developed above. 
Note that the tadpole diagram contains only the bare four-gluon vertex and is 
independent of the external momentum.\footnote{The tadpole diagram can even be 
completely absorbed in the renormalization.} 
For the other diagrams we obtain the following degree of IR divergence:
\beq
(a)  \; \rho = -2\kappa , \quad 
(b)  \; \rho =   \kappa , \quad 
(c)  \; \rho = 2 \kappa , \quad 
(d)  \; \rho = 2 \kappa .  
\eeq
Clearly, the ghost-loop diagram is the leading diagram in the IR, and the
analysis leading to eqs.\ (\ref{Zbeh}) is thus not altered.

With analogous techniques one can straightforwardly prove that the generic
IR behaviour of a $2n$-ghost--$m$-gluon amputated and
connected Greens function is given by $(p^2)^{(n-m)\kappa}$.

To summarize this part, assuming that skeleton expansions of higher
$n$-point Greens functions are well-defined we have shown:\\
$\bullet$ The degree of nonperturbative IR divergence of an $n$-point
function is given already by the highest degree of divergence present in
the lowest order skeleton expansion of the diagrams appearing in its DSE.\\
$\bullet$ In Landau gauge Yang-Mills theory the three-gluon vertex
behaves as $(p^2)^{-3\kappa}$ and the four-gluon vertex as 
$(p^2)^{-4\kappa}$ in the IR.

A consistency check on this IR behaviour is, of course, given by its
implication on the IR value of the running coupling as infered from 
these vertex functions. As we will see, these results allow for
universality of the IR fixed point in Yang-Mills theory.
A nonperturbative expression for the running coupling has been derived 
in the context of
DSEs from the ghost-gluon vertex \cite{vonSmekal:1997is}. The
Slavnov-Taylor identity (STI)
\begin{equation} \label{norm-sti}
\widetilde{Z}_{1} = Z_g\: \widetilde{Z}_3\: Z_3^{1/2} , 
\end{equation}
relates the vertex renormalization factor $\widetilde{Z}_{1}$ with the
corresponding factor $Z_g$ for the coupling $g$ and the ghost and gluon
fields, $\widetilde{Z}_3^{1/2}$, $Z_3^{1/2}$. With the definition 
$\alpha = g^2/(4\pi)$ and the relation $g(\Lambda^2) = Z_g(\mu^2,
\Lambda^2)g(\mu^2)$ one obtains 
\begin{equation} \label{1}
\alpha(\mu^2) = \alpha(\Lambda^2)  \frac{\widetilde{Z}_3^2(\mu^2, 
\Lambda^2)  Z_3(\mu^2, \Lambda^2)}{\widetilde{Z}_{1}^2(\mu^2,
\Lambda^2)} 
\end{equation}
for the renormalized coupling at the renormalization point $\mu^2$. The
bare coupling $\alpha(\Lambda^2)$
depends on a regulator $\Lambda^2$. Since the ghost-gluon
vertex is finite in Landau gauge, one may choose $\widetilde{Z}_{1}=1$.
The bare and renormalized ghost and gluon dressing functions are related by
\beq
G_0(p^2,\Lambda^2) = G(p^2,\mu^2) \widetilde{Z}_3(\mu^2,\Lambda^2) \; , 
\quad 
Z_0(p^2,\Lambda^2) = Z(p^2,\mu^2) {Z}_3(\mu^2,\Lambda^2) \; .
\eeq
Substituting these relations into eq.(\ref{1}) we obtain
\begin{equation} \label{2}
\alpha(\mu^2)  G^2(p^2,\mu^2) Z(p^2,\mu^2)  = \alpha(\Lambda^2)
G_0^2(p^2,\Lambda^2) Z_0(p^2,\Lambda^2) \; .
\end{equation}
Note that the r.h.s.\ is independent of $\mu^2$ and thus the
l.h.s.\ is renormalization group invariant \cite{Mandelstam:1979xd}.
Evaluating the l.h.s.\  once 
at an arbitrary renormalization point $\mu^2$ and once at $\mu^2=p^2$ we
obtain
\begin{equation} \label{coupling}
\alpha(p^2)  = \alpha(\mu^2)  G^2(p^2,\mu^2)  Z(p^2,\mu^2)
\end{equation}
where we have exploited the renormalization condition $G^2(p^2,p^2) 
Z(p^2,p^2)=1$  for the renormalized dressing functions. 
Evaluating $\alpha(p^2)$ in the ultraviolet  one recovers the well known
perturbative coupling in the MOM-scheme which can be related 
to the $\overline{MS}$-coupling by standard techniques  
\cite{Celmaster:1979km}.

The IR behaviour of this coupling can be read off from the power
law behaviour of the ghost and 
gluon dressing functions in eqs.\ (\ref{Zbeh}). Due to the
interrelated exponents in these expressions
the IR momentum dependence cancels and leads to a fixed point at $p^2=0$.
The precise value of $\alpha(0)$ depends on $\kappa$. It
has been calculated 
in both, DSEs and ERGEs, employing several kinds of truncation schemes 
\cite{vonSmekal:1997is,Atkinson:1998tu,Zwanziger:2001kw,Lerche:2002ep,Fischer:2002hn,Pawlowski:2003hq,Fischer:2004uk}.
Neglecting IR corrections from the dressing of the ghost-gluon vertex 
yields 
\begin{equation}
\alpha(0) = \frac{2 \pi}{3 \, N_c}  \frac{\Gamma(3-2\kappa) 
\Gamma(3+\kappa)  \Gamma(1+\kappa)}
{\Gamma^2(2-\kappa)  \Gamma(2\kappa)} \approx  8.915/N_c \; .
\end{equation}

For the running coupling based on the three-gluon vertex we employ
the STI
\begin{equation}\label{sti2}
{Z}_{1} = Z_g\: Z_3^{3/2}
\end{equation}
relating the renormalization factor of the three-gluon vertex to the one
of the coupling and the gluon fields.
The vertex renormalization factor ${Z}_{1}$ relates the regularized and
the renormalized three-gluon vertex
via
\begin{equation}
\Gamma^{\mu \nu \lambda }(p_1,p_2,p_3,\mu^2) = \Gamma_0^{\mu \nu
\lambda}(p_1,p_2,p_3,\Lambda^2) \:\:Z_1(\mu^2,\Lambda^2) \; .
\end{equation}
Evaluating the vertex at the symmetric point $p_1^2 = p_2^2=p_3^2 \equiv
p^2$ one obtains 
\begin{equation}
H_1^{3g}(p^2,\mu^2) = H_1^{3g}(p^2,\Lambda^2) \:\:Z_1(\mu^2,\Lambda^2)
\end{equation}
and thus a relation between the renormalized and the regularized $H_1^{3g}$
as defined in eq.\ (\ref{3g-symmvert}). 
One can also derive an expression for the three-gluon coupling:
\begin{equation} \label{3g-coupling}
\alpha^{3g}(p^2)  = \alpha^{3g}_0(\mu^2) \: (H_1^{3g})^2(p^2,\mu^2) \:  
Z^3(p^2,\mu^2) 
\end{equation}
where we exploited the renormalization condition $(H_1^{3g})^2(p^2,p^2) \:
Z^3(p^2,p^2)=1$. This condition involves only $H_1^{3g}$ because it multiplies the
tensor structure of the three-gluon vertex containing the primitive UV
divergence.  Again the r.h.s.\ of this equation denotes an RG invariant
quantity. This coupling is related to the one of the ghost-gluon vertex 
by a known scale transformation \cite{Celmaster:1979km}. 

Recalling that $H_1^{3g}(p^2) \sim (p^2)^{-3\kappa}$ and $Z(p^2) \sim
(p^2)^{2\kappa}$
we obtain
\begin{equation} \label{3g-coupling-fix}
\alpha^{3g}(p^2 \rightarrow 0)  \sim  {const.}/{N_c}  \; ,
\end{equation}
{\it i.e.\/} the running coupling taken from the three-gluon vertex 
possesses an IR  fixed point in accordance with the coupling determined
from the ghost-gluon vertex.
The explicit $1/N_c$-dependence stems from the fact that $g^2 \sim 1/N_c$
and the functions $H_1^{3g}$ and $Z$ are independendent of $N_c$, as can be
seen from their DSEs. 
Note that 
 also other renormalization conditions for
 the three-gluon vertex involving $H_2^{3g}$ and $H_3^{3g}$ are 
 possible \cite{Celmaster:1979km}. The expression
 (\ref{3g-coupling}) for the coupling would then
 include a corresponding linear combination of the functions $H_1^{3g}$, $H_2^{3g}$
 and $H_3^{3g}$. All of these functions are proportional to $(p^2)^{-3\kappa}$
 in the IR, therefore such a redefinition 
 of the coupling would not affect the
 presence of the IR fixed point.  

Finally we examine the four-gluon coupling. Its STI is
${Z}_{4} = Z_g^2\: Z_3^{2},$
and thus we obtain the relation
\begin{equation}
H^{4g}_1(p^2,\mu^2) = H^{4g}_1(p^2,\Lambda^2) \:\:Z_4(\mu^2,\Lambda^2) \; .
\end{equation}
The running coupling from the four-gluon vertex is therefore given by  
\begin{equation} \label{4g-oupling}
\alpha^{4g}(p^2)  = \alpha^{4g}_0(\mu^2) \:H^{4g}_1(p^2,\mu^2) \:
Z^2(p^2,\mu^2) 
\end{equation}
where analogously to the three-gluon vertex case 
the renormalization condition $H^{4g}_1(p^2,p^2)
\: Z^2(p^2,p^2)=1$ has been employed. 
Recalling that $H^{4g}_1(p^2) \sim (p^2)^{-4\kappa}$ and $Z(p^2) \sim
(p^2)^{2\kappa}$ in the IR we  again obtain an IR fixed point,
\beq 
\alpha^{4g}(p^2 \rightarrow 0)  \sim  {const.}/{N_c}  .
\eeq

To summarize, we have demonstrated on a qualitative level that the running
couplings from the ghost-gluon, three-gluon and four-gluon vertices
have a universal, nontrivial fixed point in the IR. The value of the fixed 
point for the ghost-gluon coupling is known to depend slightly on the IR 
dressing of the ghost-gluon vertex and falls into a small window 
$2.5 < \alpha(0) < 3$ for $N_c=3$ \cite{Lerche:2002ep}. 
The corresponding value for the other couplings can be 
calculated from the vertex-DSEs. In the skeleton expansion the dressed 
three- and four-gluon vertices receive their IR leading contributions from 
an infinite number of diagrams belonging to a certain subclass. These can 
be constructed from the ghost-triangle diagram by repeated insertion of 
the diagrammatic pieces given in fig.\ \ref{skelkorr2}. In the IR, all 
these diagrams have the same dependence on one external momentum scale. 
In principle, the momentum independent coefficients can be calculated order
by order in the skeleton expansion, although its convergence properties are 
yet to be determined.
The nonperturbative expressions for the three- and four-gluon vertices are
IR singular provided all external squared momenta approach 
zero. This singularity, however, is not strong enough
to compensate the zeroes in the gluon dressing functions of attached gluon 
propagators.
Thus in the skeleton expansion of all DSEs the ghost-loop diagrams
are leading in the IR. This is in agreement with a picture recently
advocated  by Zwanziger \cite{Zwanziger:2003cf}: the geometric degrees of
freedom, {\it i.e.\/} the Faddeev-Popov determinant,  dominate IR Yang-Mills 
theory. 

{\emph{We are grateful to Holger Gies, Lorenz von Smekal, Mike Pennington, Peter Watson
and Dan Zwanziger for helpful discussions.\\ 
This work has been supported by a
grant from the Ministry of Science, Research and the Arts of
Baden-W\"urttemberg (Az: 24-7532.23-19-18/1 and 24-7532.23-19-18/2),
the Deutsche Forschungsgemeinschaft (DFG) under contract Fi 970/2-1,
and Spanish MCYT FPA 2004-02602, BFM 2002-01003.}}



\begin{thebibliography}{99}

\vspace*{-6mm}

{\footnotesize
\setlength{\baselineskip}{12pt plus 1pt}

\parskip=2pt
\bibitem{Bonnet:2001uh}
F.~D.~Bonnet {\it et al.},
Phys.\ Rev.\ D {\bf 64}  (2001) 034501.

\bibitem{Langfeld:2001cz}
K.~Langfeld, H.~Reinhardt and J.~Gattnar,
Nucl. Phys. B {\bf 621} (2002) 131.

\bibitem{Bowman:2004jm}
P.~O.~Bowman, U.~M.~Heller, D.~B.~Leinweber, M.~B.~Parappilly and A.~G.~Williams,
Phys.\ Rev.\ D {\bf 70} (2004) 034509.

\bibitem{Silva:2004bv}
P.~J.~Silva and O.~Oliveira,
Nucl.\ Phys.\ B {\bf 690} (2004) 177.

\bibitem{Alkofer:2001wg}
R.~Alkofer and L.~von Smekal,
Phys.\ Rept.\  {\bf 353} (2001) 281.

\bibitem{vonSmekal:1997is}
L.~von Smekal, R.~Alkofer and A.~Hauck,
Phys.\ Rev.\ Lett.\  {\bf 79}  (1997) 3591;
L.~von Smekal, A.~Hauck and R.~Alkofer,
Annals Phys.\  {\bf 267} (1998) 1.

\bibitem{Atkinson:1998tu}
D.~Atkinson and J.~C.~Bloch,
Mod.\ Phys.\ Lett.\ A {\bf 13} (1998) 1055; \newline
Phys.\ Rev.\ D {\bf 58} (1998) 094036.


\bibitem{Zwanziger:2001kw}
D.~Zwanziger,
Phys.\ Rev.\ D {\bf 65} (2002)  094039.

\bibitem{Lerche:2002ep}
C.~Lerche and L.~von Smekal,
Phys.\ Rev.\ D {\bf 65} (2002) 125006.

\bibitem{Fischer:2002hn}
C.~S.~Fischer  and R.~Alkofer,
Phys. Lett. {\bf B536}  (2002) 177; \newline
C.~S.~Fischer, R.~Alkofer and H.~Reinhardt,
Phys.\ Rev.\ {\bf D65}  {2002} 094008; \newline
R.~Alkofer, C.~S.~Fischer and L.~von Smekal,
Acta Phys.\ Slov.\  {\bf 52} (2002)  191; \newline
C.~S.~Fischer,
PhD thesis, U.\ of Tuebingen, arXiv:hep-ph/0304233.

\bibitem{Alkofer:2003jj}
R.~Alkofer, W.~Detmold, C.~S.~Fischer and P.~Maris,
Phys.\ Rev.\ D {\bf 70} (2004) 014014; 
arXiv:hep-ph/0411367.
\bibitem{Kugo:1995km}
T.~Kugo,
Int.\ Symp.\ on BRS symmetry, Kyoto, Sep.~18-22, 1995, arXiv:hep-th/9511033.

\bibitem{Zwanziger:2003cf}
D.~Zwanziger,
Phys.\ Rev.\ D {\bf 69} (2004) 016002.

\bibitem{Celmaster:1979km}
W.~Celmaster and R.~J.~Gonsalves,
Phys.\ Rev.\ D {\bf 20} (1979) 1420.

\bibitem{Shirkov:1997wi}
D.~V.~Shirkov and I.~L.~Solovtsov,
Phys.\ Rev.\ Lett.\  {\bf 79} (1997) 1209.

\bibitem{Solovtsov:1999in}
I.~L.~Solovtsov and D.~V.~Shirkov,
Theor.\ Math.\ Phys.\  {\bf 120} (1999) 1220
[Teor.\ Mat.\ Fiz.\  {\bf 120} (1999) 482].

\bibitem{Howe:2003mp}
D.~M.~Howe and C.~J.~Maxwell,
Phys.\ Rev.\ D {\bf 70} (2004) 014002.

\bibitem{Pawlowski:2003hq}
J.~M.~Pawlowski, D.~F.~Litim, S.~Nedelko and L.~von Smekal,
Phys.\ Rev.\ Lett.\  {\bf 93} (2004) 152002.


\bibitem{Fischer:2004uk}
C.~S.~Fischer and H.~Gies,
JHEP 0410 (2004) 048.

\bibitem{Stevenson:1981vj}
P.~M.~Stevenson,
Phys.\ Rev.\ D {\bf 23} (1981) 2916.

\bibitem{Brodsky:1982gc}
S.~J.~Brodsky, G.~P.~Lepage and P.~B.~Mackenzie,
Phys.\ Rev.\ D {\bf 28} (1983) 228.

\bibitem{cotanch} F. J. Llanes-Estrada and S. R. Cotanch, 
arxiv:nucl-th/0408038; \newline
F. J. Llanes-Estrada {\it et al.\/},
Phys. Rev. C{\bf 70} (2004) 035202.

\bibitem{Zwanziger:1990by}
D.~Zwanziger,
Phys.\ Lett.\ B {\bf 257} (1991) 168.

\bibitem{Taylor:ff}
J.~C.~Taylor,
Nucl.\ Phys.\ B {\bf 33} (1971) 436.

\bibitem{Marciano:1977su}
W.~J.~Marciano and H.~Pagels,
Phys.\ Rept.\  {\bf 36} (1978) 137.


\bibitem{Schleifenbaum:2004id}
W.~Schleifenbaum, A.~Maas, J.~Wambach and R.~Alkofer,
arXiv:hep-ph/0411052.

\bibitem{Cucchieri:2004sq}
A.~Cucchieri, T.~Mendes and A.~Mihara,
arXiv:hep-lat/0408034.


\bibitem{Watson:2001yv}
P.~Watson and R.~Alkofer,
Phys.\ Rev.\ Lett.\  {\bf 86} (2001) 5239.

\bibitem{Mandelstam:1979xd}
S.~Mandelstam,
Phys.\ Rev.\ D {\bf 20} (1979) 3223.

\bibitem{Boos:1990rg}
E.~E.~Boos and A.~I.~Davydychev,
Theor.\ Math.\ Phys.\  {\bf 89} (1991) 1052
[Teor.\ Mat.\ Fiz.\  {\bf 89} (1991) 56].

\bibitem{Anastasiou:1999ui}
C.~Anastasiou, E.~W.~N.~Glover and C.~Oleari,
Nucl.\ Phys.\ B {\bf 572} (2000) 307.

\bibitem{Watson:2000}
P.~Watson, Ph.D thesis, University of Durham, UK, 2000.
%
}
\end{thebibliography}
\end{document}